\documentclass[10pt, doublecolumn]{IEEEtran}
\usepackage{epsfig,latexsym}
\usepackage{float}
\usepackage{indentfirst}
\usepackage{amsmath}
\usepackage{bm}
\usepackage{amssymb}
\usepackage{times}

\usepackage{algorithm}
\usepackage[noend]{algpseudocode}
\usepackage{subfigure}
\usepackage{psfrag}
\usepackage{hyperref}
\usepackage{cite}
\usepackage{lastpage}
\usepackage{fancyhdr}
\usepackage{color}
 \usepackage{amsthm}
\usepackage{bigints}
\newcounter{problem}
\newcounter{save@equation}
\newcounter{save@problem}
\makeatletter
\newenvironment{problem}
{\setcounter{problem}{\value{save@problem}}%
  \setcounter{save@equation}{\value{equation}}%
  \let\c@equation\c@problem
  \subequations
}
{\endsubequations
  \setcounter{save@problem}{\value{equation}}%
  \setcounter{equation}{\value{save@equation}}%
}

\begin{document}
\title{  \vspace{-0.5em}\huge{   A Simple Study on the Optimality of  Hybrid NOMA     }}

\author{ Zhiguo Ding, \IEEEmembership{Fellow, IEEE}       \thanks{ 
  
\vspace{-2em}

Z. Ding is with Khalifa University, Abu Dhabi, UAE, and University
of Manchester, Manchester, M1 9BB, UK.   
 

  }\vspace{-2.5em}}
 \maketitle

\begin{abstract}
The key idea of hybrid non-orthogonal multiple access (NOMA) is to allow users to use the bandwidth resources to which they cannot have access in orthogonal multiple access (OMA) based legacy networks while still guaranteeing its compatibility with the legacy network. However, in a conventional hybrid NOMA network,   some users have access to more bandwidth resources than others, which leads to a potential performance loss. So what if the users can access the same amount of bandwidth resources? This letter focuses on a simple two-user scenario, and develops analytical and simulation results to reveal that for this considered scenario, conventional hybrid NOMA is still an optimal transmission strategy.     
\end{abstract}\vspace{-0.5em}

\vspace{-0.5em} 

\section{Introduction}
As a potential candidate of the next-generation multiple access,    hybrid non-orthogonal multiple access (NOMA)  not only supports spectrum sharing, the key principle of NOMA, but also ensures that its implementation is transparent to   conventional  orthogonal multiple access (OMA) based networks, where its superior compatibility feature yields an effective coexistence between OMA and NOMA \cite{Scizigu}.  
 
\begin{figure}[t!]\centering \vspace{-0em}
    \epsfig{file=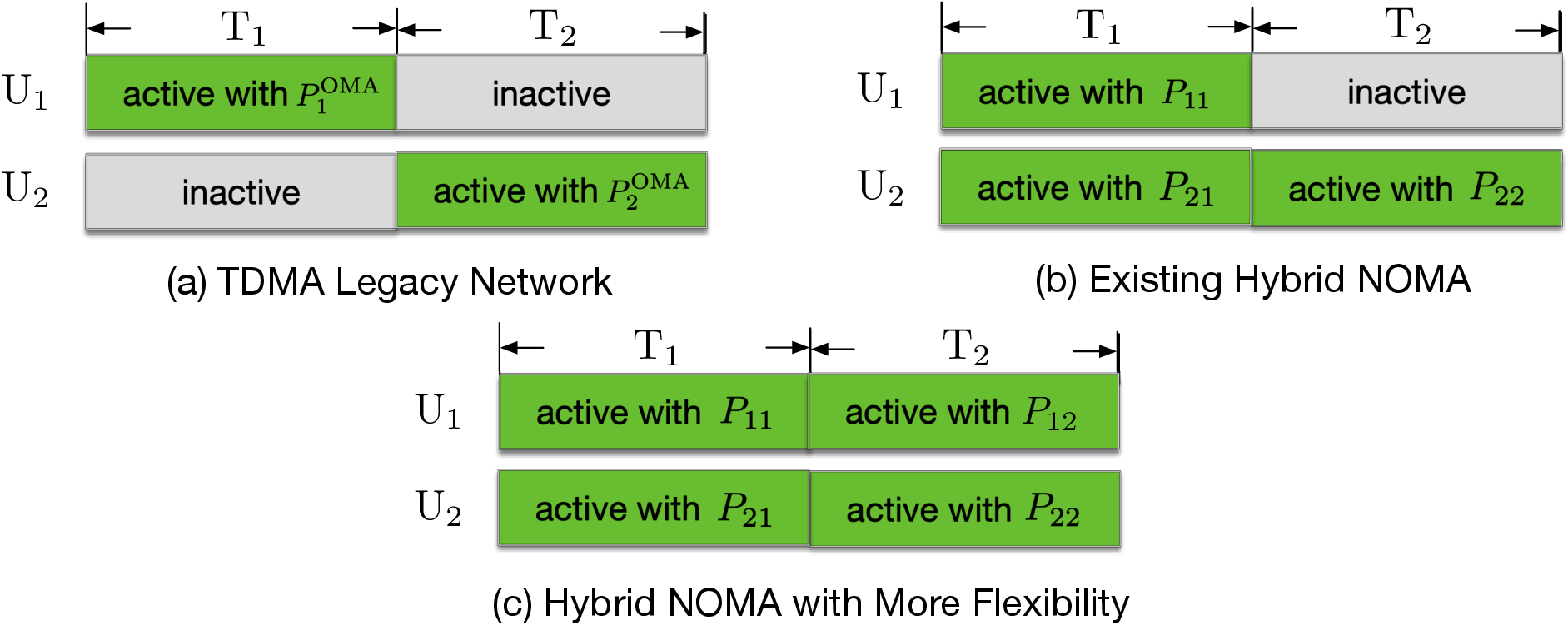, width=0.4\textwidth, clip=}\vspace{-0.5em}
\caption{Illustration for the three considered transmission strategies.     \vspace{-2em}    }\label{fig0}   \vspace{-1.5em} 
\end{figure}

The key idea of hybrid NOMA can be illustrated by the two-user scenario shown in Fig. \ref{fig0}. Assume that there exists a   time division multiple access (TDMA) based legacy network, as shown in Fig. 1(a). Using hybrid NOMA can ensure that one user has access to the time slot which belongs to another user in the legacy network, as shown in Fig. 1(b). As shown in \cite{Scizigu, 10470406,hnomadown,10379617}, the use of hybrid NOMA can significantly reduce energy consumption and improve the system throughput, compared to the OMA case.  One feature of hybrid NOMA is that some users have access to more bandwidth resources than others, which is to ensure the compatibility with OMA, and simplify the system design. However, this feature can lead to a potential performance loss, which is the motivation of this work. In particular, this letter is to investigate an extension of hybrid NOMA, as shown in Fig. 1(c), where each user is offered the same access to the available bandwidth resources. The question that this letter seeks to answer is whether the system performance can be improved with this extension.   Analytical results are developed to reveal that the optimal transmission strategy for the case shown in Fig. 1(c)  achieves the same performance as the conventional hybrid NOMA strategy shown in Fig. 1(b). In other words, for the considered two-user case, conventional hybrid NOMA is an optimal transmission strategy, even if the users are provided with the same access to the available bandwidth resources.   
   
\vspace{-0.5em}
\section{System Model} 
Assume that there exists a TDMA based legacy downlink network, where two users,  denoted by ${\rm U}_m$, $m\in \{1,2\}$,  are served by the same base station. Without loss of generality, assume that the $m$-th time slot, denoted by ${\rm T}_m$, is assigned to ${\rm U}_m$ in the legacy network, and ${\rm U}_m$'s channel gain is denoted by   $g_{m}$,  where quasi-static fading is assumed, i.e., ${\rm U}_m$'s channel gains over the two time slots are same.  To be consistent with the hybrid NOMA literature  \cite{hnomadown}, it is assumed that $h_1\geq h_2$, where $h_m=|g_m|^2$. 
 
It is straightforward to show that with OMA, the base station can serve ${\rm U}_m$ with the following transmit power: $
 P^{\rm OMA}_m = \frac{e^R-1}{h_m}$,
 where it is assumed that the two users have the same target data rate, denoted by $R$. 
 \vspace{-1em}
 \subsection{Conventional   Hybrid NOMA }
 As shown in Fig. \ref{fig0}, the key idea of the conventional hybrid NOMA strategy is to offer ${\rm U}_2$ access to the additional time slot, i.e., ${\rm T}_1$. Denote $P_{m,n}$ by the transmit power for ${\rm U}_m$'s signal at ${\rm T}_n$.  By following the same steps shown in \cite{hnomadown}, it is straightforward to show that hybrid NOMA outperforms OMA, and the transmit powers for ${\rm U}_2$'s signals in the two time slots are given by
 \begin{align}
P_{2,1}^{\rm HN} = \sqrt{\frac{e^R}{h_2^2 \mu^{\rm HN}}} - \frac{1}{\mu^{\rm HN} h_2},   P_{2,2}^{\rm HN}  = \sqrt{\frac{e^R}{h_2^2 \mu^{\rm HN}}} - \frac{1}{h_2},
\end{align}
where   $\mu^{\rm HN}=\frac{1}{P^{\rm HN} _{1,1}h_2+1}$ and $P^{\rm HN} _{1,1}=\frac{e^{R}-1}{h_1}$. Note that $h_1\geq h_2$  guarantees $P_{2,1}^{\rm HN}  \geq 0$.  
\vspace{-1em}
\subsection{An Extension of  Hybrid NOMA}
We note that hybrid NOMA does not help ${\rm U}_1$, whose transmit powers in OMA and hybrid NOMA are the same. The rationale behind this design is that in the legacy network,  ${\rm U}_1$ should already finish its transmission by the end of  ${\rm T}_1$, and hence there is no need to offer ${\rm U}_1$   access to ${\rm T}_2$. Otherwise, an additional latency could be introduced at ${\rm U}_1$.

This letter is to investigate whether the system performance can be further enhanced, if   ${\rm U}_1$ is delay-tolerant and can also use  ${\rm T}_2$, as illustrated in Fig. 1(c). As to be shown in the next section, conventional hybrid NOMA surprisingly turns out to be still optimal for this considered scenario.  In other words, ${\rm U}_1$ does not use the additional time slot (${\rm T}_2$), even if it is offered access to this time slot. 
 
 \vspace{-0.5em}
\section{Optimality of   Hybrid NOMA} \label{section optimality}
The fact that there are two users to be served at each time slot makes the design of successive interference cancellation (SIC) more complex.     In order to ensure compatibility with the legacy network, at  ${\rm T}_m$,  only ${\rm U}_m$, which is the legacy user at this time slot, carries out SIC,   i.e., it first decodes its partner's signal before decoding its own. Denote $R_{i,j}$ by ${\rm U}_i$'s achievable data rate at ${\rm T}_j$,  Therefore, at ${\rm T}_1$, the two users' achievable data rates are given by
{\small \begin{align}\label{rates in ts1}
R_{1,1} = \log\left(
1+P_{1,1}h_1
\right),\quad R_{2,1} = \log\left(
1+\frac{P_{2,1}h_2}{P_{1,1}h_2+1}
\right),
\end{align}}
\hspace{-0.4em}where the noise power is assumed to be normalized. We note that due to the use of SIC, a more rigorous expression of $R_{2,1} $ should be $\min \left\{
\log\left(
1+\frac{P_{2,1}h_1}{P_{1,1}h_1+1}
\right), \log\left(
1+\frac{P_{2,1}h_2}{P_{1,1}h_2+1}
\right)
\right\}$, where the simplified expression in \eqref{rates in ts1} is obtained by applying the assumption that $h_1\geq h_2$. The two users' data rates at ${\rm T}_2$ can be obtained in a similar way:
{\small \begin{align}\label{rates in ts2}
R_{1,2} = \log\left(
1+\frac{P_{1,2}h_2}{P_{2,2}h_2+1}
\right),\quad R_{2,2} = \log\left(
1+P_{2,2}h_2
\right),
\end{align}}
\hspace{-0.4em}where $R_{1,2}=\min \left\{
\log\left(
1+\frac{P_{1,2}h_1}{P_{2,2}h_1+1}
\right),\log\left(
1+\frac{P_{1,2}h_2}{P_{2,2}h_2+1}
\right)
\right\}$ can be simplified as in \eqref{rates in ts2} due to the fact that $h_1\geq h_2$. 
As can be seen from \eqref{rates in ts1} and \eqref{rates in ts2}, a legacy user's data rate at its own time slot is exactly the same as that in the OMA mode, which ensures the compatibility with the OMA based legacy network. It is also important to point out that ${\rm U}_1$'s data rate at ${\rm T}_2$ is decided by ${\rm U}_2$'s channel gain, $h_2$, instead of $h_1$. 

Similar to \cite{hnomadown}, the considered resource allocation problem is to minimize the total power consumption:
  \begin{problem}\label{pb:1} 
  \begin{alignat}{2}
\underset{P_{m,n}\geq0   }{\rm{min}} &\quad   \sum^{2}_{m=1}  \sum^{2}_{n=1}P_{m,n}\label{1tst:1}
\\ s.t. &\quad  \sum^{2}_{n=1} R_{m,n}\geq R, \quad 1\leq m \leq 2.  \label{1tst:2}  
  \end{alignat}
\end{problem}

By assuming that ${\rm U}_1$'s parameters are fixed,   problem \eqref{pb:1} can be recast as the following optimization problem with respect to ${\rm U}_2$'s parameters:  
  \begin{problem}\label{pb:2} 
  \begin{alignat}{2}
\underset{P_{2,n}\geq0   }{\rm{min}} &\quad   P_{2,1}+P_{2,2}\label{1tst:1}
\\ s.t. &\quad   \log\left(
1+\frac{P_{2,1}h_2}{P_{1,1}h_2+1}
\right) + \log\left(
1+P_{2,2}h_2
\right)\geq R.  \label{1tst:2}  
  \end{alignat}
\end{problem} 
We note that   problem \eqref{pb:2} is convex,  and  by following the same steps shown in \cite{hnomadown}, its optimal solution can be obtained as follows: 
\begin{align}\label{optimal fake}
P_{2,1}^* = \lambda - \frac{1}{\mu h_2}, \quad P_{2,2}^* = \lambda - \frac{1}{h_2},
\end{align}
i.e., ${\rm U}_2$ chooses the hybrid NOMA mode, where $\lambda = \sqrt{\frac{e^R}{h_2^2 \mu}}$ and  $\mu=\frac{1}{P_{1,1}h_2+1}$, if $P_{1,1}h_2+1\leq e^R$. Otherwise $P_{2,1}^* = 0$, and $ P_{2,2}^* = \frac{e^R-1}{h_2}
$, i.e., ${\rm U}_2$ chooses the pure  OMA mode.

By substituting the closed-form expressions of $P_{2,1}^* $ and $ P_{2,2}^*$  in problem \eqref{pb:1},   an equivalent form of the original optimization problem can be obtained as follows:
  \begin{problem}\label{pb:3} 
  \begin{alignat}{2}
\underset{P_{1,n}\geq0   }{\rm{min}} &\quad   P_{1,1}+P_{1,2}\label{1tst:1}
\\ s.t. &\quad  \log\left(
1+P_{1,1}h_1
\right)+  \log\left(
1+\frac{P_{1,2}h_2}{P_{2,2}^*h_2+1}
\right) \geq R.  \label{1tst:2}  
  \end{alignat}
\end{problem}

\subsection{${\rm U}_2$ Chooses the Pure OMA Mode}
In this case, $P_{2,1}^* = 0$ and $ P_{2,2}^* = \frac{e^R-1}{h_2}
$. Hence, problem \eqref{pb:3} can be rewritten as follows:
  \begin{problem}\label{pb:4} 
  \begin{alignat}{2}
\underset{x,y\geq0   }{\rm{min}} &\quad   x+y\label{1tst:1}
\\ s.t. &\quad  \log\left(
1+xh_1
\right)+  \log\left(
1+ e^{-R}h_2y
\right) \geq R,  \label{1tst:2}  
  \end{alignat}
\end{problem}
where $x=P_{1,1}$ and $y=P_{1,2}$. It is straightforward to verify that problem \eqref{pb:3} is a convex optimization problem. Its corresponding Lagrangian can be expressed as follows:
\begin{align}\nonumber
L=&x+y +\lambda_1\left(R- \log\left(
1+xh_1
\right)-  \log\left(
1+ e^{-R}h_2y
\right)\right)\\  
&-\lambda_2x-\lambda_3y,
\end{align}
where $\lambda_i$ denotes the Lagrange multipliers.  
By using the above Lagrangian, the  Karush–Kuhn–Tucker (KKT) conditions of problem \eqref{pb:3} can be obtained as follows:  \cite{Boyd}
\begin{align}
\left\{\begin{array}{l}  
1 -\lambda_1  \frac{h_1}{1+xh_1}  -\lambda_2 =0\\
1 -\lambda_1 \frac{e^{-R}h_2}{1+ e^{-R}h_2y}  -\lambda_3=0
\\
\lambda_1\left(R- \log\left(
1+xh_1
\right)-  \log\left(
1+ e^{-R}h_2y
\right)\right)=0\\
 \lambda_2x=0,  \lambda_3y=0
 \end{array}\right..
 \end{align}

In order to show  the optimality of hybrid NOMA, it is sufficient to show that the OMA solution,  i.e., $x=\frac{e^R-1}{h_1}$ and $y=0$, satisfies all the KKT conditions, since for a convex optimization problem, the KKT conditions are the sufficient and necessary conditions for a solution to be optimal.  
 
Because $x=\frac{e^R-1}{h_1}$, $\lambda_2=0$.  By using the expression of the OMA solution and the fact that $\lambda_2=0$, the KKT conditions can be simplified  as follows:
 \begin{align}
\left\{\begin{array}{l}  
1 -\lambda_1  h_1e^{-R}  =0\\
1 -\lambda_1 e^{-R}h_2  -\lambda_3=0 
 \end{array}\right. \Rightarrow \left\{\begin{array}{l}  
 \lambda_1   = \frac{ e^{R}}{h_1}\\
\lambda_3=1 -\frac{ h_2}{h_1}      
 \end{array}\right. .
 \end{align}
 Because it is assumed that $h_1\geq h_2$, $\lambda_3\geq 0$, which means that the OMA solution is an optimal solution of problem \eqref{pb:4}. Therefore, if ${\rm U}_2$ chooses the OMA mode, ${\rm U}_1$ also chooses the OMA mode.

\begin{figure*}\vspace{-2em}
\begin{align} \label{figureeq}
\nabla^2 f(z,y) =  \begin{bmatrix}
-\frac{1}{(z-a)^2}
+
\frac{3}{4}  \frac{e^{-\frac{1}{2}R} h_2 ^{\frac{1}{2}} y  z^{-\frac{5}{2}}}{1+  e^{-\frac{1}{2}R} h_2 ^{\frac{1}{2}} y  z^{-\frac{1}{2}}}
-
\frac{1}{4}  \frac{e^{- R} h_2   y^2  z^{-3}}{\left(1+  e^{-\frac{1}{2}R} h_2 ^{\frac{1}{2}} y  z^{-\frac{1}{2}}\right)^2}
&\frac{1}{z-a}-\frac{1}{2}  \frac{e^{-\frac{1}{2}R} h_2 ^{\frac{1}{2}} y  z^{-\frac{3}{2}}}{1+  e^{-\frac{1}{2}R} h_2 ^{\frac{1}{2}} y  z^{-\frac{1}{2}}}
\\
\frac{e^{-\frac{1}{2}R} h_2 ^{\frac{1}{2}}  z^{-\frac{1}{2}}}{1+  e^{-\frac{1}{2}R} h_2 ^{\frac{1}{2}} y  z^{-\frac{1}{2}}}&\frac{e^{-\frac{1}{2}R} h_2 ^{\frac{1}{2}}  z^{-\frac{1}{2}}}{1+  e^{-\frac{1}{2}R} h_2 ^{\frac{1}{2}} y  z^{-\frac{1}{2}}}
\end{bmatrix}
\end{align}
\end{figure*}
\subsection{${\rm U}_2$ Chooses the Hybrid NOMA Mode}
For this case, the expressions in \eqref{optimal fake} can be used, and problem \eqref{pb:3} can be recast as follows:
 \begin{problem}\label{pb:5} 
  \begin{alignat}{2}
\underset{x,y\geq0   }{\rm{min}} &\quad   x+y\label{1tst:5}
\\ s.t. &\quad   \log\left(
1+xh_1
\right)+  \log\left(
1+\frac{ e^{-\frac{1}{2}R} h_2  y }{\sqrt{ xh_2+1}}
\right) \geq R \label{2tst:5}
\\  & \quad x h_2+1\leq e^R,
  \end{alignat}
\end{problem}
where the constraint, $x h_2+1\leq e^R$, is needed as it is the condition for the solution in \eqref{optimal fake} to be chosen. 

It is interesting to point out that problem \eqref{pb:5} is not always a convex optimization problem. In particular, its constraint function in \eqref{2tst:5} can be expressed as follows: $\tilde{f}_1(x,y)\triangleq \log(h_1)+\tilde{f}_2(z,y)$, where $z=x+\frac{1}{h_2}$,   $
\tilde{f}_2(z,y)=
 \log\left(
z -a
\right)+  \log\left(
1+  e^{-\frac{1}{2}R} h_2 ^{\frac{1}{2}} y  z^{-\frac{1}{2}}
\right) $, and $a=\frac{1}{h_2}-\frac{1}{h_1}$. With some algebraic manipulations, the Hessian matrix of $\tilde{f}_2(z,y)$ can be obtained as shown in \eqref{figureeq} at the top of the next page. Computer simulations show that $\nabla^2 f(z,y) $ is not always a positive semidefinite matrix, i.e., its convexity depends on the choices of $h_1$, $h_2$, $R$, $y$ and $z$.

Instead of directly solving this non-convex optimization problem shown in \eqref{pb:5}, by fixing $x$, the optimal solution of $y$ can be straightforwardly obtained as follows:   
\begin{align} 
    y^* 
  = \frac{e^{\frac{3}{2}R}}{h_2}\frac{\sqrt{ xh_2+1}}{1+xh_1} -\frac{e^{\frac{1}{2}R}}{h_2}\sqrt{ xh_2+1}, 
\end{align}
if $\frac{e^{\frac{3}{2}R}}{h_2}\frac{\sqrt{ xh_2+1}}{1+xh_1} -\frac{e^{\frac{1}{2}R}}{h_2}\sqrt{ xh_2+1}\geq 0$. Otherwise, the considered optimization problem becomes a problem without a feasible solution.   Note that the considered condition  can be simplified as follows: 
\begin{align}
    \frac{e^{\frac{3}{2}R}}{h_2}\frac{\sqrt{ xh_2+1}}{1+xh_1} -\frac{e^{\frac{1}{2}R}}{h_2}\sqrt{ xh_2+1}\geq 0 \rightarrow 1+xh_1\leq e^R.
\end{align}

Therefore,    problem \eqref{pb:5} can be simplified as the following optimization problem with respect to $x$ only: 
 \begin{problem}\label{pb:6} 
  \begin{alignat}{2}
\underset{P_{m,n}\geq0   }{\rm{min}} &\quad  f(x)\triangleq  x+\frac{e^{\frac{3}{2}R}}{h_2}\frac{\sqrt{ xh_2+1}}{1+xh_1} -\frac{e^{\frac{1}{2}R}}{h_2}\sqrt{ xh_2+1}\label{1tst:1}
\\ s.t. &\quad  x h_1+1\leq e^R,
  \end{alignat}
\end{problem}
where the assumption $h_1\geq h_2$ is applied to remove the constraint $x h_2+1\leq e^R$. Problem \eqref{pb:6} is a convex optimization problem, as shown in the following. 

The first and second order derivatives of $f(x)$ can be obtained as follows:
\begin{align}\label{first order}
f'(x) =& 1+\frac{e^{\frac{3}{2}R}}{2 }\frac{ \left( xh_2+1\right)^{-\frac{1}{2}}}{1+xh_1}-\frac{e^{\frac{3}{2}R}h_1}{h_2}\frac{\left( xh_2+1\right)^{\frac{1}{2}}}{(1+xh_1)^2}\\\nonumber & -\frac{e^{\frac{1}{2}R}}{2}\left( xh_2+1\right)^{-\frac{1}{2}}
,\\\nonumber 
f''(x) =& -\frac{e^{\frac{3}{2}R}h_2}{4 }\frac{ \left( xh_2+1\right)^{-\frac{3}{2}}}{1+xh_1}
- \frac{e^{\frac{3}{2}R}h_1}{2 }\frac{ \left( xh_2+1\right)^{-\frac{1}{2}}}{(1+xh_1)^2}\\\nonumber & 
-\frac{e^{\frac{3}{2}R}h_1}{2 }\frac{\left( xh_2+1\right)^{-\frac{1}{2}}}{(1+xh_1)^2}
+\frac{2e^{\frac{3}{2}R}h_1^2}{h_2}\frac{\left( xh_2+1\right)^{\frac{1}{2}}}{(1+xh_1)^3}
\\\nonumber & +\frac{e^{\frac{1}{2}R}h_2}{4}\left( xh_2+1\right)^{-\frac{3}{2}}.
\end{align}

It is challenging to directly show $f''(x)\geq 0$. First define $\phi(x)=4h_2(1+xh_1)^3 \left( xh_2+1\right)^{\frac{3}{2}}e^{-\frac{1}{2}R}f''(x)$. The first order derivative of $\phi(x)$ is given by

\begin{align}  
\phi'(x)=& - 6e^{ R}h_1h_2^2  
+  6e^{ R}h_1^2 h_2^2 x 
  \\\nonumber &  +12e^{ R}h_1^2 h_2
+ 3h_1h_2^2  (1+xh_1)^2\geq 0,
\end{align}
where the inequality follows by the assumption that $h_1\geq h_2$. Therefore, $\phi(x)$ is a monotonically increasing function of $x$. Furthermore, note that $\phi(0)$ is given by
\begin{align}
\phi(0)=&  \left( e^{ R}h_1^2  - e^{ R}h_2^2\right)   +
\left(4e^{ R}h_1^2-  4e^{ R}h_1 h_2    \right)\\\nonumber &
+ 3e^{ R}h_1^2  
+ h_2^2   \geq 0,
\end{align}
which means that $\phi(x)\geq \phi(0)\geq 0$. Therefore,  $f''(x)\geq 0$, and   problem \eqref{pb:6} is indeed a convex optimization problem. 
%

By using the first order derivative expression shown in \eqref{first order}, the KKT conditions of  problem \eqref{pb:6} can be expressed
as follows:
\begin{align}
\left\{\begin{array}{l}  
1+\frac{e^{\frac{3}{2}R}}{2 }\frac{ \left( xh_2+1\right)^{-\frac{1}{2}}}{1+xh_1}-\frac{e^{\frac{3}{2}R}h_1}{h_2}\frac{\left( xh_2+1\right)^{\frac{1}{2}}}{(1+xh_1)^2} \\ -\frac{e^{\frac{1}{2}R}}{2}\left( xh_2+1\right)^{-\frac{1}{2}}+\lambda_4 h_1-\lambda_5=0
\\
\lambda_4\left(x h_1+1- e^R
\right) =0\\
 \lambda_5x=0 
 \end{array}\right.,
 \end{align}
where $\lambda_4$ and $\lambda_5$ are the Lagrange multipliers. 

 Again, because of the convexity of the considered optimization problem, these KKT conditions can be used as the certificate   to show the optimality of OMA. With OMA, $x=\frac{e^R-1}{h_1}$, which means that  $\lambda_5=0$ and $x h_1+1=e^R$. Therefore, the KKT conditions can be simplified as follows: 
\begin{align}
 1  -\frac{e^{-\frac{1}{2}R}h_1}{h_2}\left( \frac{e^R-1}{h_1} h_2+1\right)^{\frac{1}{2}} +\lambda_4 h_1=0,
\end{align}
which means that $\lambda_4$ can be obtained as follows: 
\begin{align} 
\lambda_4 = \frac{e^{-\frac{1}{2}R} }{h_2}\left( \frac{e^R-1}{h_1} h_2+1\right)^{\frac{1}{2}} -\frac{1}{h_1}. 
\end{align}
To show the optimality of OMA, it is important to show that $\lambda_4$ is non-negative.  
We note that $\lambda_4\geq 0$ is equivalent to the following inequality:  
\begin{align}
  1  \leq\frac{e^{-\frac{1}{2}R}h_1}{h_2}\left( \frac{e^R-1}{h_1} h_2+1\right)^{\frac{1}{2}}  ,
  \end{align}
  which can be further simplified as follows:
  \begin{align} 
 e^Rh_2^2   \leq e^Rh_1h_2 +h_1^2-h_1h_2.
\end{align}
The above inequality holds, since $h_1\geq h_2$. This means that   the OMA solution is optimal for problem \eqref{pb:6}. Therefore, if ${\rm U}_2$ chooses the hybrid NOMA mode, ${\rm U}_1$ still chooses the OMA mode. 
 
 In summary, ${\rm U}_1$ always chooses the OMA mode, regardless of the choices made by ${\rm U}_2$. In other words, ${\rm U}_1$ does not use ${\rm T}_2$, even if it has access to this time slot, which means that hybrid NOMA is optimal for the two-user case. 

%
%
  
     \begin{figure}[t!]\centering \vspace{-2em}
    \epsfig{file=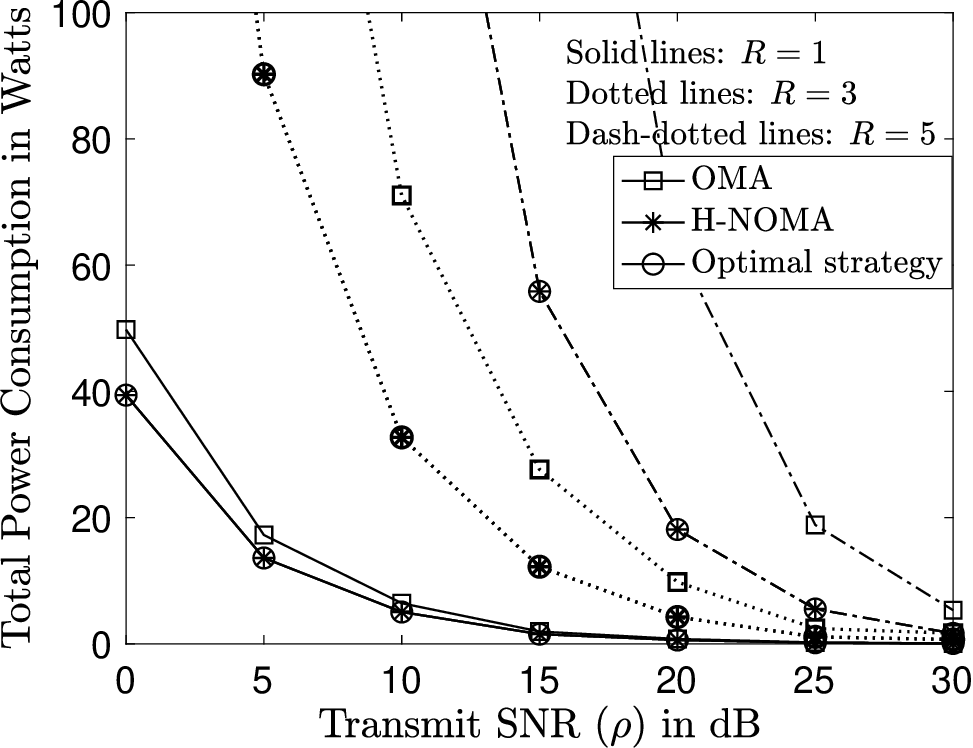, width=0.3\textwidth, clip=}\vspace{-0.5em}
\caption{Total power consumption required by the three considered transmission strategies as a function of the transmit signal-to-noise ratio (SNR), denoted by $\rho$,  where the order users' channel gains are obtained from two independent and identically complex Gaussian distributed random variables with zero mean and variance $\rho$.    \vspace{-1em}    }\label{fig1}   \vspace{-0.2em} 
\end{figure}
       \begin{figure}[t!]\centering \vspace{-0em}
    \epsfig{file=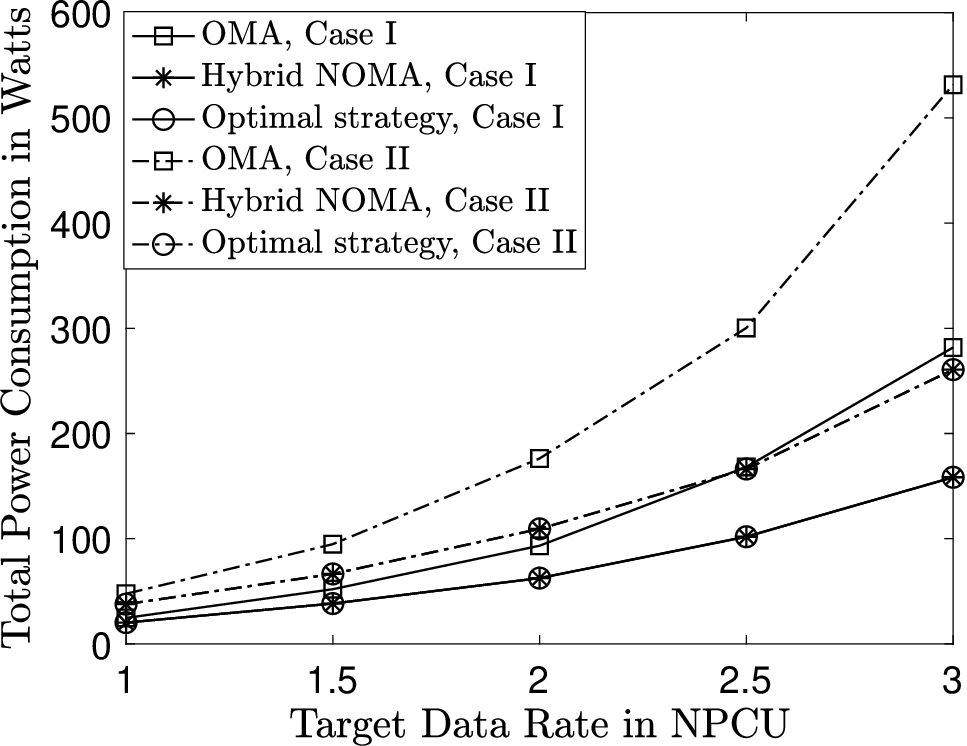, width=0.3\textwidth, clip=}\vspace{-0.5em}
\caption{Total power consumption required by the three considered transmission strategies as a function of the target data rate in nats per channel use (NPCU). With Case I, the ordered users' channels are obtained from two independent and identically complex Gaussian distributed random variables with zero mean and unit variance.   With Case II, the variances of two random variables are $\frac{1}{2}$ and   $1$, respectively.   \vspace{-2em}    }\label{fig2}   \vspace{-0.2em} 
\end{figure}
\section{Numerical Results}
In this section, the computer simulation results are presented to verify the optimality of hybrid NOMA in the considered downlink scenario, where each user has the same access to the two time slots.  

In Fig. \ref{fig1}, the total power consumption required by the optimal transmission strategy of the considered communication scenario, i.e., the optimal solution of problem \eqref{pb:1}, is shown as a function of the transmit signal-to-noise ratio (SNR), where OMA and hybrid NOMA are used as the benchmarking schemes. We note that problem \eqref{pb:1} can be decomposed into two optimization problems, namely problems \eqref{pb:4} and \eqref{pb:6}, and the analysis in Section \ref{section optimality}  shows that the solution of problem \eqref{pb:6} outperforms that of problem \eqref{pb:4}. Therefore, the curve for problem \eqref{pb:1} is obtained by applying an optimization solver to problem \eqref{pb:6}.   As can be seen from the figure, the optimal transmission strategy for the considered scenario and hybrid NOMA achieve the same performance, which verifies the optimality of hybrid NOMA. In Fig. \ref{fig2}, the total power consumption required by the three considered transmission strategies is shown as a function of $R$. Similar to Fig. \ref{fig1}, the optimal transmission strategy yields the same performance as hybrid NOMA, which again verifies the optimality of hybrid NOMA.  We note that the performance gain of hybrid NOMA over OMA in Case II is larger than that of Case I, which means that it is more beneficial to use hybrid NOMA if the users' channel difference is larger. 

     \begin{figure}[t] \vspace{-2em}
\begin{center}
\subfigure[Channel Realization I]{\label{fig6a}\includegraphics[width=0.3\textwidth]{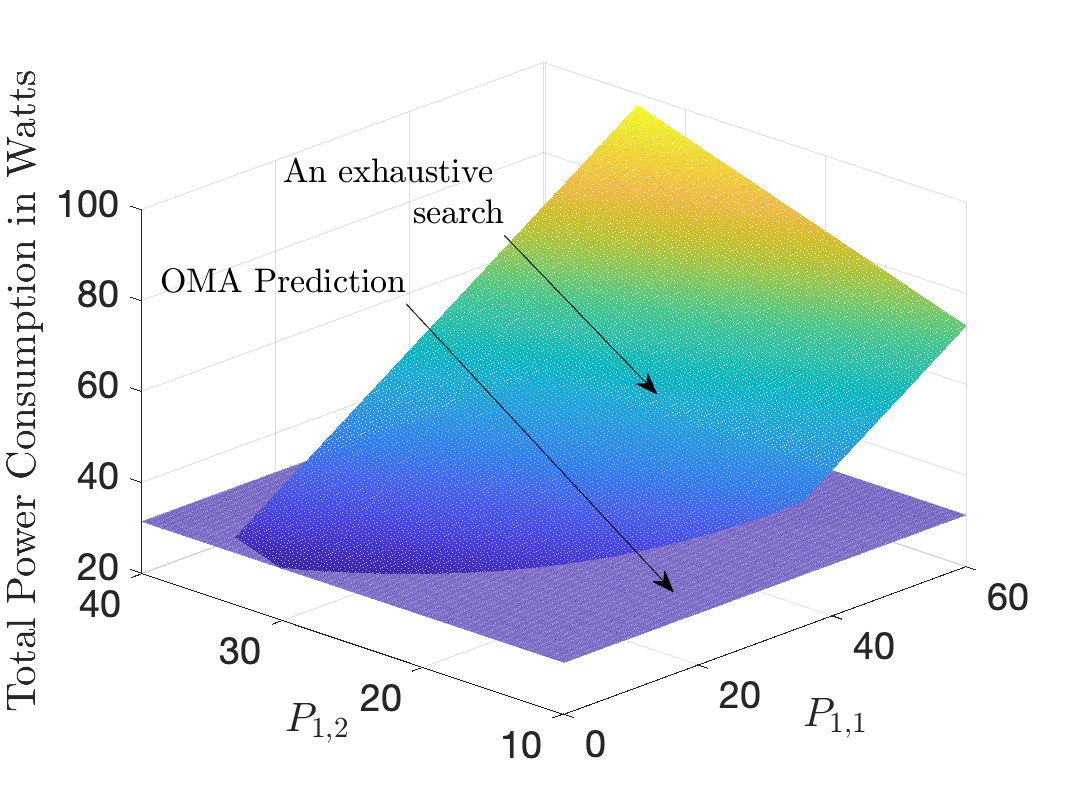}} 
\subfigure[ Channel Realization II]{\label{fig6b}\includegraphics[width=0.3\textwidth]{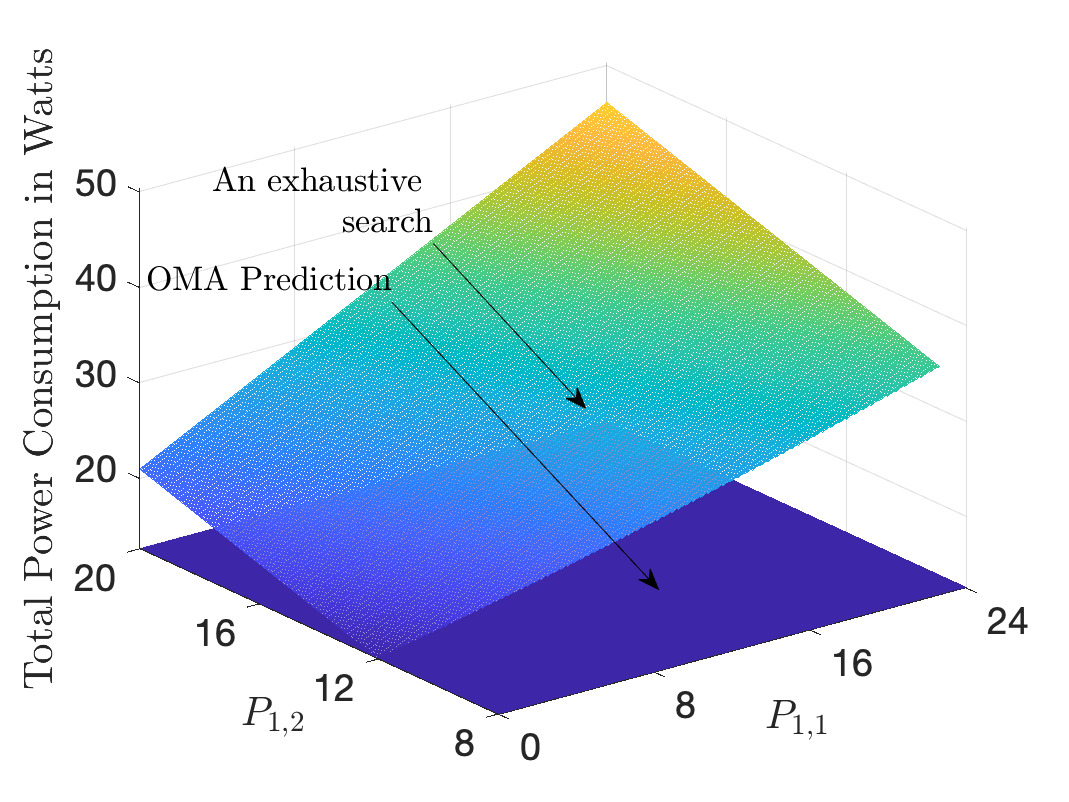}} \vspace{-1em}
\end{center}
\caption{Analysis of the optimal solution of problem \eqref{pb:3}, where $R=3$ NPCU. In the two subfigures, two random realizations are used for the users' channels, by assuming that the ordered users' channels are obtained from two independent and identically complex Gaussian distributed random variables with zero mean and unit variance.   \vspace{-1em} }\label{fig3}\vspace{-1em}
\end{figure}

Recall that one key question to be answered in this letter is whether ${\rm U}_1$ uses ${\rm T}_2$, if it is offered access to  ${\rm T}_2$. The answer to this question can be revealed by analyzing the solution of problem \eqref{pb:3}. In particular, if the solution of problem \eqref{pb:3} is different from the OMA solution,  ${\rm U}_1$ can reduce its power consumption by choosing a strategy other than OMA, which means that hybrid NOMA is not optimal. In Fig. \ref{fig3},   an exhaustive search is used to find the optimal solution of problem \eqref{pb:3}. As can be seen from Fig. \ref{fig3}, the optimal solution of problem \eqref{pb:3} achieves the same performance as OMA. In other words, ${\rm U}_1$'s optimal strategy is still to use OMA, even if  the access to the extra time slot, ${\rm T}_2$, is offered to  ${\rm U}_1$.  
\vspace{-1em}
\section{Conclusions}
In this letter, for a simple two-user downlink scenario,   analytical and simulation results were provided to reveal that conventional hybrid NOMA is still an optimal transmission strategy. However, for more complex scenarios, e.g., cases with more than two users and time-varying channels, the optimality of hybrid NOMA needs to be further investigated, which is an important direction for future research.     
\vspace{-0.5em}
\bibliographystyle{IEEEtran}
\bibliography{IEEEfull,trasfer}
  \end{document}